# Analysis of the Independent Particle Model approach to Nuclear Densities

F. B. Guimaraes

*Instituto de Estudos Avançados/DCTA,*

*12228-001 São José dos Campos, São Paulo, Brazil*

*e-mail: fbraga@ieav.cta.br*

## Abstract

*We present an analysis of the use of the Darwin-Fowler approximation in connection with the statistical IPM, by comparing the results of our recent studies with the occupation number approach (OCN) and some tradtional statistical independent particle model (IPM) approaches. The analysis of level density works based on the statistical IPM reveals that the use of the the Darwin-Fowler approximation, in some of them, is theoretically inconsistent and some of their results should rather be considered as coincidences with other consistent approaches, than proofs of their validity. We conclude that, in general, the use of the Darwin-Fowler approximation with the statistical IPM should be used criteriously or, if possible, avoided altogether and suggest that the combinatorial IPM approaches have important advantages over the other models and formalisms analyzed in this paper, especially regarding the consistency of the microscopic description of the nuclear structure and dynamics of non highly excited systems.*

## 1. Introduction

Our recent study of the nuclear level density and the second moments of the nuclear Hamiltonian in the pre-equilibrium stage of nuclear reactions (PE),[1, 2] has shown that some problems result in connection with the Darwin-Fowler method (DFM),[3] due to the statistical nature of its approximations. It lacks the precision to completely describe the details of the microscopic interaction of the PE processes for the evaluation of the moments, in connection with the use of Laplace transform.[1]

The method developed in Ref.[3] considers a set of Planck vibrators (PV) that may be excited to a series of *discrete* energies and does not necessarily define an approximate continuum of single particle levels. The method is centered on the approximate evaluation of integrals over a statistical variable that is initially only an auxiliary parameter used to define the possible "complexions" of the assembly (in our case, the nuclear states). Later this variable is identified with the thermodynamic function $e^{-\beta\epsilon}$, where $\beta$ is the inverse of the nuclear temperature and $\epsilon$ is the elementary energy defining the modes of vibration.

We have seen in Ref.[4] and it was also stressed by Darwin and Fowler [3], that a necessary condition for the applicability of the saddle point or "steepest descent" approximation is that the energy of the "complexions" is *large enough*. We have also shown that the method is not applicable in general if the various thermodynamic parameters are considered complex at the same time and, for non highly excited nuclear systems, the method should be avoided.[4]

Having in sight the generality of the Darwin-Fowler formalim, due to the use of the Cauchy theorem with possible ramification into the Laplace transform (by using the continuous approximation for the nuclear levels), this suggests that there may be a problem of consistency in nuclear models, connected with the use of the DFM.

On the other hand, various works,[5, 6, 7, 8] have criticized the statistical approach to the highly degenerated Fermi gas model[9] as a method to describe the nuclear level density, in part due to the small number of particles of the nuclear system, but also due to constraints imposed by the approach on the excitation energy, $U$, and the lack of sensibility to the model adopted for the single particle levels. Due to the statistical nature of the Fermi gas model, which may suggest the use of DFM, the analysis of these works could bring some clarification on the existence of the above mentioned problem of consistency.

In this context, it is possible to identify, in general, three basic approaches that have been proposed to obtain the nuclear level density. The first, which we will call simply the "statistical approach" (STA), is based on the seminal work of Bethe[9] in which the level density is defined by systematic approximations to the higly degenerate Fermi gas model and by the expressions for the number of particles, $A$, and excitation energy, $U$, from the original "theory of metals" of Sommerfeld.[10] The influence of the nuclear angular momentum is carefully crafted into the model, using the grand canonical definition of the thermodynamic potential and approximated integrals, to obtain general expressions for the nuclear level density that can be used as a guide for the construction of phenomenological models.

The "highly degenerate" hypothesis in the model is equivalent to the assumption that the nuclear excitation, $U$, is small in comparison with the sum of energies of all single particle nucleon states (total nuclear kinetic energy) and that the formulas from the theory of metals are applicable. The nucleon states are supposed to be given by a simple model, typically the wave-functions of free particles in a spherical box, and the continuous approximation ($CAP$) for single particle levels is supposed valid, i. e. the hypothesis that the single particle states are so closely packed together that it is reasonable to replace the sums over state energies by the corresponding integrals modulated by the single particle state density.

Ref.[9] also uses the hypothesis that the nuclear temperature is close to zero and consistenply *avoids* the use of the Darwin-Fowler method, which makes the proposed statistical formalism physically coherent and powerful despite its approximate character. This traditional approach has served as basis for many successful phenomenological level density models like the Gilbert-Cameron[11] and the Mengoni-Nakagima[12] descriptions, but the connection with the microscopic details of the single particle levels is blurred by the systematic approximations and the replacement of the sums over single particle levels by integrals.

A second formulation for the nuclear density, which we will call the "occupation number approach" (OCN),[6] is the one in which the basic model equations define the *occupation of single particle states* as a continuous function of the nuclear excitation to compose the various possible nuclear configurations. In this case, the density of single particle levels is not supposed to be large enough to define a "continuum", as implicitly assumed in the STA, but a connection between the two models can be made using the definition of temperature of the STA for the analysis of the level density as a function of the atomic mass and nuclear excitation.[6]

A third approach is based on the *direct distribution of the nucleons* into a set of predefined single particle levels, with[13, 14, 15, 16] or without[5] the direct counting of degeneracies of nuclear levels associated with the spin-isospin symmetry.

In the works of Ref.[13, 14, 15, 16] the concept of "nuclear configuration" is that of a given, *fixed*, set of single particle levels and the corresponding number of assumed nucleons per level. Their approach is used mainly in connection with the description of the nuclear levels of ground states and low lying excited states of specific nuclei. It is closely related with the analysis of the degeneracy of nuclear levels associated with the spin-isospin symmetry. (see Appendix I)

We use a similar direct distribution approach,[1, 2, 4] but define the "nuclear configuration" differently, as *any nuclear state* that can be obtained by combinatorial means by the distribution of nucleons into a set of predefined single particle states, constrained only by the exclusion princicple and the conservation laws. The "Model Space" is defined as usual, as the specific subsets of the nucleons and single particle states of each nuclear configuration that are excitable ("mobile") or not, the definition of the ground state energy and configuration, the adopted model for single particle levels and the number of hole states.

All these approaches assume an approximate model for the nucleons, in which they are supposed to be moving nearly independently of each other in a common mean field, which is usually designated as the "independent particle model" (IPM).

In the direct distribution approach one considers the degeneracy of the nuclear system as defined by the distribution of nucleons itself and by the "symmetries" associated with the maximum occupation of single particle levels, governed by the exclusion principle applied to spin, isospin, orbital angular momentum and other additive quantum numbers. We will call this direct approach the "combinatorial independent particle model" (IPMC) and it can be considered as a generalization of the OCN, in which the total occupations of the nuclear levels are not continuous functions of the nuclear temperature and, therefore, more consistent with the quantum description.

The Darwin-Fowler method should be applied only to systems with high excitation[3, 4] and, in this regard, some of these works may already be classified. One can see that the DFM was used inconsistently in Refs.[6, 7], without specification of the excitation energy range of validity, and it was used consistently in Ref.[8], by adopting a different approach (direct counting) for low energies. There are also examples in the literature of the IPMC being used either in connection with the Darwin-Fowler method[3, 17, 18] or not,[5] but none of these works analyze in detail the possible weaknesses of the use of the DFM to improve the classical STA approach.[9] Other formalisms, like the one of Ref.[8], can be considered as a kind of "hybrid model" between STA and IPMC.

In this paper we add some points of criticism to those presented in Ref.[4] to show that the connection of the IPM with the statistical Darwin-Fowler approximation is problematic. We compare our results with the OCN and some tradtional statistical IPM approaches[7, 17, 18] , except the STA, and present the arguments of why the use of the IPMC to define the level density should be favored instead.

We will not analyze in detail the STA approach of Ref.[9] as we consider it as a basic prescription to serve as a starting point for phenomenological approximations, rather than a fundamental description of the nuclear system.

The present work does not intend to bring new results for the calculation of nuclear densities or to present a new formalism, but to compare previous works with respect to the use of the DFM in the various approaches defined above (and the related consistency problems) and to compare them with the IPMC.[1, 4, 5]

Using this criterion, we analyze in detail the method of Refs. [7] and [17] in Sec.2, the OCN approach in section Sec.3 and present our final comments and conclusions in Sec.4. For completeness and also to serve as reference, we present a description of the direct counting of degeneracies in connection with the spin-isospin symmetry and the STY-parameters of Wigner in Appendix I. This traditional formalism is completely independent of the Darwin-Fowler method and can be considered as part of the IPMC.

## 2.  Modified statistical IPM using the saddle-point approximation

The inadequacy of the use of a continuous distribution to approximate the sequence of single particle levels has been remarked in the literature[5, 7, 8] in connection with attempts to derive a more precise formulation for the nuclear level scheme than the STA approach.[9]

Refs.[5, 7] give the argument of large interspacing between levels, in the case of low mass nuclei, to indicate that the approximation of the discrete single particle levels by a *continuum* is not generally valid, but Ref.[7] still relies on the Darwin-Fowler method to obtain the final expressions of the proposed nuclear level model and also uses the "highly degenerate Fermi gas" assumption to describe the nucleus, in which only the levels close to the top of the Fermi distribution are important to define the Model Space.

The "high degeneracy" implies a very large number of states within a relatively small range of excitations, which would permit the use of *CAP* and the replacement of sums over levels by the corresponding integrals, with the low excitation implying low average kinetic energy of the component particles and warranting their nearly independent movement, as in a *gas*. This is a coherent approach only inasmuch as the approximations of the DFM are not introduced, as they imply high excitations.[3, 4]

The Darwin-Fowler method was not considered in the statistical approach of Bethe[9], which, therefore, is more reliable for the description of low energy processes than the formalism proposed by Bloch[7, 17]. The existence of inconsistencies in Bloch's formalism, due to the use of DFM with no clear specification of the range of validity of the method, makes it less reliable, theoretically speaking, despite the interesting results obtained in Refs.[7, 17].

On the other hand, the IPMC estimate of nuclear configurations confirms, in general, the high degeneracy of highly excited nuclear levels, even if the spin-isospin symmetries of the nuclear structure are not taken into account to define the available nuclear states,[1] but, depending on the single particle model used, the excited levels may *not* be very close to the Fermi level and usually the nuclear degeneracy for low nuclear excitation is not large, specially for light nuclei.[19]

Therefore, the restriction of the analysis to levels close to the Fermi level and low temparatures, as in the formalism of Ref.[7], may not be reasonable if one expects to use *CAP* consistently to replace the sums over nuclear levels by integrals, as the DFM presupposes. In general, the statistical IPM modified by the Darwin-Fowler method is not expected to give good estimates of level densities in all ranges of excitations, and different approaches should be used to describe the low energy region.[8]

The necessity of a different approach for low energies may also be true for the STA, because in this region the cumulative number of states may not be very large, specially for medium and light nuclei.[11]. Nonetheless, independently of the number of states for low energies, the level density is still well defined and the energy quantization of the levels can be described exactly using Dirac functions. Therefore, the existence or not of high degeneracy of the single particle levels does not necessarily affect the definition of the nuclear density, and does not affect the theoretical consistency of the STA model.

The formalism of Ref.[7] and some results are analyzed in greater detail in the next section.

## 2.1  A sophisticate but inconsistent description of level densities

Bloch's approach[7] introduced a new algrabraic formalism to define nuclear densities, in which the systematic approximations of the STA could be avoided, to some extent, and the details of the single particle level model adopted could be more precisely taken into account. One of the basic assumptions of the proposed formalism, though, was the exactness of the Darwin-Fowler approach to describe the nuclear level density, as suggested by the original formulation of Ref.[3].

In the Darwin-Fowler formalism the level density, $\rho(A,E)$, can be defined as the pole of the grand canonical generating function, $f(x,y)$, which can be written as [4]

$$f(x,y) = \prod_i (1 + xy^{\nu_i}) \ ,$$

(2.1)

divided by adequate factors $x^{A+1}y^{E+1}$,

$$\rho(A,E) = \frac{1}{(2\pi i)^2 \epsilon} \oint \oint \frac{f(x,y)dxdy}{x^{A+1}y^{N+1}} \ ,$$

(2.2)

where $x$ and $y$ are parameters associated with the chemical potential, $\mu$, and the nuclear temperature, $T$: $x=e^{\beta\mu}$, $y=e^{-\beta\epsilon}$, with $\beta=1/\kappa T$, and $\kappa$ is the Boltzmann constant.

Either for fermions, Eq. (2.1), or for bosons[4] the grand canonical partition function can be expanded as a sum of products of terms of the type $(xy^{\nu_i\epsilon})$, where $\nu_i$ is an integer, $\epsilon$ is an adequate energy unit, and the coefficients of the expansion are given by integrals like (2.2). In fact, except for the physical meaning of the parameteres $x$ and $y$, these integrals actually define an *exact formal solution* for the coefficients of the expansion[3]. Notice that higher moments of the nuclear Hamiltonian can also be defined similarly.[1, 2]

For the Darwin-Fowler method to be physically meaningful though, the terms of the expansion must decrease in modulus when $y$ (or $x$) vary over complex circles around the origin in comparison with its value at the positive real axis. This point of maximum must also be a minimum along the positive real axis, thus defining a *saddle point* located on the positive real axis of $y$ (or $x$).[3, 17] In addition, the maximum along the contour should be "strong" to permit to use the contribution to the integral in the neighborhood of that point as a good approximation for the entire integral, so that the terms that do not show strong oscillation in this neighborhood can be "factorated" out of the integral sign and the other contributions neglected.

Another important concept used in the formalism of Ref.[7] is that of a single particle "component" of the grand canonical nuclear ensemble, which is equivalent, for description of the ensemble, to our concept of nuclear configuration, as we explain next with an example.

At the center of the traditional statistical IPM description of the nucleus as a Fermi gas[6, 7, 9] is the Sommerfeld approximation for the thermodynamic potential[10]

$$\phi_\alpha = \beta(\epsilon_\alpha N_\alpha - W_\alpha) + (\pi^2/6\beta)\rho_\alpha\rho_\alpha\epsilon_\alpha) \ , \tag{2.3}$$

where $\beta$ is the inverse of the nuclear temperature. Eq.(2.3) can be deduced[17] on the assumption that $\beta$ is very large (low excitations) and that $\epsilon_F = \mu$ is the Fermi level. Then, $N_\alpha$ is given by

$$N_\alpha = \int_0^{\epsilon_\alpha} \rho_\alpha(\epsilon)d\epsilon \tag{2.4}$$

and $W_\alpha$ is

$$W_\alpha = \int_0^{\epsilon_\alpha} \epsilon\rho_\alpha(\epsilon)d\epsilon \tag{2.5}$$

where $\rho_\alpha(\epsilon)$ is the density of single particle energies for the "component $\alpha$" of the total set of nucleon states, $\mathcal{N}$, and for a "highly degenerate system" the important values of $\beta$ are supposed to be the "large ones", therefore the temperature should be close to zero and $\rho_\alpha$ should fall quickly for $\epsilon > \epsilon_\alpha$. Here $\epsilon_\alpha$ is the Fermi energy of the component $\alpha$.

The components of $\mathcal{N}$ are classified in Ref.[7] according to their values of the constants of motion, $c_k$. If the set $\{c_k; k = 0, 1, \cdots, K\}$ is complete then, in the case of fermions, each fermion state $\alpha$ will be associated with one set of values of the constants of motion $\mathcal{C}_\alpha = \{\mathtt{m_k}; k = 0, 1, \cdots, K\}_\alpha$, where "$\mathtt{m_0}$" is always "1", corresponding to the powers of the parameter "$x$" in the grand partition function (2.1). If each fermion state is completely characterized by energy and total angular momentum only then, $K = 1$ and $\mathcal{C}_\alpha = \{\mathtt{m_0}, \mathtt{m_1}\}_\alpha$. Each different nucleon state is then characterized by a set $\mathcal{C}_\alpha$ that Bloch calls a "component" of the total description of single particle states, or a "component" of $\mathcal{C}$, where

$$\mathcal{C} = \{\mathcal{C}_\alpha;\ \alpha \in \{\text{all possible sets of good quantum numbers for nucleons}\}\} \ . \tag{2.6}$$

Then, for example, if the fermions can be characterized by single particle energy and spin (two quantum numbers) and the number of single particle energies is $N_{\mathbf{e}}$ and the number of spins is 2 ("up" and "down"), the total number of components will be $2N_{\mathbf{e}}$. Each component of $\mathcal{C}$ is one single particle state that *may be* occupied by one or more "nucleons" (or "holes") to define part of a nuclear configuration and will be characterized, in this example, by a subset like

$$\mathcal{G}(i,\alpha) = \{\epsilon_i, 1, s_\alpha\} \ . \tag{2.7}$$

Notice that the notation of Ref.[7] is a little imprecise because the energies of the single particle levels, $\epsilon_i$, belong to the set of the constants of motion of the system of fermions, but they are not considered along with the other constants to define a "component" (of the set of single particle states).

Then, in the more general case of $K$ constants of motion Eq.(2.7) becomes

$$\mathcal{G}(i,\alpha) = \{\epsilon_i, \mathbf{m}_{\alpha \mathbf{k}}; k \in \{0, \cdots, K-2\}\} \quad ; \text{ with } \ \epsilon_i = \epsilon \nu_i \ , \ \nu_i = \text{ integer} \tag{2.8}$$

For example, in the case of 4 constants of motion, including energy, the fermionic partition function can be written as

$$f(x,y,w,s) = \prod_{i,j,k}(1 + xy^{\nu_i}w^{\mathbf{m}_\mathbf{j}}s^{n_k}) = \tag{2.9}$$

$$= 1 + x\sum_{(i,j,k)} y^{\nu_i}w^{\mathbf{m}_\mathbf{j}}s^{n_k} + x^2 \sum_{(i_1,i_2)}\sum_{(j1,j2)}\sum_{(k1,k2)} y^{\nu_{i1}+\nu_{i2}}w^{\mathbf{m}_{\mathbf{j}1}+\mathbf{m}_{\mathbf{j}2}}s^{n_{k1}+n_{k2}}$$

$$\cdots + x^A \sum_{(i_1,\cdots,i_A)}\sum_{(j1,\cdots,jA)}\sum_{(k1,\cdots,kA)} y^{\nu}w^{M}s^{N} \tag{2.10}$$

where,

$$\nu = \nu_{i1} + \cdots + \nu_{iA} = (\epsilon_{i1} + \cdots + \epsilon_{iA})/\epsilon \tag{2.11}$$

$$M = \mathbf{m}_{\mathbf{j}1} + \cdots + \mathbf{m}_{\mathbf{j}A} \tag{2.12}$$

$$N = n_{k1} + \cdots + n_{kA} \tag{2.13}$$

therefore, the partition function $f(x,y,w,s)$ can be described as a sum over all possible single particle "components", like $\{\alpha\}_\alpha = \{(i_{\mathtt{k}_1}, \mathtt{m}_{\mathtt{k}_2}, n_{\mathtt{k}_3})\}_{\mathtt{k}_1, \mathtt{k}_2, \mathtt{k}_3}$, where $\mathtt{k}_1$, $\mathtt{k}_2$ and $\mathtt{k}_3$ may have an infinite range, including "hole" states, or, in the case of a nuclear system of $A$ particles, as a sum over all possible *nuclear configurations*, each composed by $A$ single particle states, defined by all possible sets $(i_1, \cdots, i_A)$, $(\mathtt{m}_1, \cdots, \mathtt{m}_A)$ and $(n_1, \cdots, n_A)$.

Therefore, this specific algebraic formulation of Ref.[7] in terms of single particle "components" is equivalent to the description in terms of nuclear configurations that we have adopted in our recent works[1, 4]. It is not connected with the assumptions of the DFM, not constrained by the range of energies in which the model is applicable and, therefore, it does not change the consistency of the model with respect to the use of the DFM at low energies.

Using these two basic formal elements, Ref.[7] then assumes that the saddle-point approximation can be used to evaluate the level density integral, (2.2), and enough precision can be achieved if the integrand is replaced by its Taylor expansion up to second order of the values of $\beta$ close to its saddle-point value.[17]

The hypothesis that the main contributions to the integral come from the neighborhood of the saddle point, can only be made if the excitation energy, $U$, is *large enough*[3, 4, 17] and at this point the formalism of Ref.[7] becomes clearly contradictory and imprecise, because on the one hand it assumes $\beta$ very large (the hypothesis $\beta \to \infty$ is an important part of the arguments used to justify the proposed algebraic formalism) and that there are not many nucleons with energy much greater than the Fermi energy, therefore that the total system possesses an excitation energy, $U$, *not very large*. On the other hand, it uses the saddle point approximation and therefore assumes that $U$ is *not too low*.[17]

This contradiction is not followed appropriately in Bloch's formalism [7, 17] and the problem with low values of $U$ is only mentioned in [17] after the deduction of Bethe's expression for the level density for one type of nucleon, using the saddle point method.

If the saddle point method cannot be applied, then the use of the Cauchy integral to evaluate the nuclear level density becomes redundant and equivalent to the direct counting of nuclear configurations by combinatorial analysis of the distribution of nucleons in the available single particle states, instead of a physically meaningful tool to obtain useful approximations. In this case, some interesting results obtained in Ref.[7], as for example the spin distribution for the nuclear states of the light nucleus model of Ref.[16], where Bloch obtains good match between his approximate estimates and the "exact" calculation using Critchfield and Oleksa data, must be considered as *coincidental* rather than a proof of the validity of the proposed formalism.

In fact, Bloch's formal results are very similar to the STA[9] and the inconsistent use of Darwin-Fowler method indicates that *this* is the cause of the good comparison with the "exact" data of Ref.[16], rather than the metod itself. This conclusion becomes even clearer if one compares Bloch's spin distribution expression with, for example, the phenomenological approach of Gilbert and Cameron[11], which is based on the STA.

Bloch argues that a more accurate treatment of the single particle level density should start with the exact definition of the thermodynamic potential given by

$$\Phi = \log(f(x, y, w, s)) \ , \tag{2.14}$$

where the explict definition for $f(x,y,w,s)$ is used, in terms of discrete set of nucleon states as in (2.9), and "$x$" is treated approximately, with the condition that $\log(x) \to \beta \epsilon_F$ when $\beta \to \infty$ (low excitations), where $\epsilon_F$ is the Fermi energy, thus obtaining the usual statstical "interpretation" for $x$.

In the "more accurate treatment" Bloch uses the expression

$$\log(x) = \beta \epsilon_n + a + \delta x_0 \ , \tag{2.15}$$

where $\epsilon_n$ is now the energy of the last (partially) occupied nucleon level and $a$ is a parameter that should be adjusted to give $\delta x_0 \to 0$ for $\beta \to \infty$. The exact expression of the thermodynamic potential, in terms of *sums* over single particle states, is then used to obtain an approximate expression for the nuclear level density as an expansion in powers of small $1/\beta$.

The saddle point approximation is used again as correct without further analysis and the fact that values of $\beta$ are usually *not* very large in actual nuclear calculations, as we will see for example in the simple OCN model analyzed in the next section, is not analyzed either.

Therefore, the analyses and proposed formalisms of Ref.[7] and similarly Ref.[17], are *clearly flawed* due to their almost strict reliance on the "exactness" of the algebraic relations based on the Darwin-Fowler method, without a more detailed attempt to quantify the results and evaluate them in comparison with the assumptions of the DFM.

Other not so sophisticate approaches[5, 6] have tried to avoid the ambiguities that may arise in the highly degenerate Fermi gas model and improve the sensibility of the calculated level density to the specific model used for single particle levels.

In particular, the study of the nuclear levels as a function of the single particle level scheme can be realized to a good extent with a rather simple, semi-phenomenological OCN approach, as in the work of Margenau[6] that we will discuss next.

## 3. The Occupation Number Approach.

In the occupation number approach (OCN) of Ref.[6] the occupation probabilities of single particle states are smooth functions of the nuclear temperature, defined by the exclusion principle and the thermodynamic properties of the grand canonical ensemble.[20] The description of the nuclear level density, $\rho(E,A)$, is based on the statistical IPM, where the degeneracies of the single particle levels, $g_i$, and their energies, $\epsilon_i$, define "$A$" and "$E$" directly as a function of the nuclear temperature and chemical potential.

The level density is defined by the approximate expressions of the highly degenerate Fermi gas for not very high nuclear excitations,[9] but a more realistic description of the nuclear system is attempted by solving directly the equations of state, without the approximate replacement of the sums over single particle levels by integrals.

For intermediate size nuclei in the fundamental state one needs to consider only nearly 10 occupied single particle levels, which makes it simple to compute $\rho(E,A)$ using direct summations over levels.[6] In addition, the formulas of Sommerfeld of the theory of metals[10] should be valid if the assumption of "extreme degeneration" of the nuclear system, considered as a Fermi gas, is correct. In this case, only a relatively small number of terms should be necessary in the summations over single particle levels, for nuclear excitations up to about 10 MeV.[9] The assumption of high degeneracy does not necessarily correspond to high excitations because the degeneracy associated with the spin-isospin symmetry can also reach very high numbers, depending on the orbital angular momentum of the single particle level being "occupied", as we show in Appendix I.[13, 14, 15, 16]

On the other hand, the Sommerfeld approximation uses the assumption of a continuous single particle level density and it is applicable only if the nuclear excitation energy is not too high.[6] Consequently, such an approximation is incompatible with the steepest descent procedure and the Darwin-Fowler method in general.[3, 4]

With this in mind let us consider, for example, the single particle levels corresponding to the basis of the spherical potential well, consisting of 10 levels up to level "3s", with maximum occupation equal to 92 particles.[6] The scheme of levels and occupations is shown in Table I.

To describe an assembly containing $A$ particles of the same type with total energy $E$ the OCN uses the following two equations of state, in which the nuclear temperature (function of the parameter $b$) and the chemical potential (function of $a$) must be determined,

$$\sum_{(i)} \frac{g_i}{1 + ab^{\epsilon_i}} = A, \qquad \text{and} \qquad \sum_{(i)} \frac{g_i \epsilon_i}{1 + ab^{\epsilon_i}} = E \ . \qquad (3.1)$$

Here $a$ and $b$ are immediately interpreted as,

$$b = \mathsf{e}^{\beta}, \ \text{ and } \ \beta = 1/\kappa T \ , \qquad (3.2)$$

where $T$ is the nuclear temperature and $\kappa$ is the Boltzmann constant,

$$a = \mathsf{e}^{-\beta\mu}, \ \text{ and } \ \mu = \text{chemical potential} \ , \qquad (3.3)$$

and the $g_i$ are the maximum occupations (considered *fixed*) of the single particle levels, with energies $\epsilon_i$.

Notice that the above definition of "nuclear system" of Ref.[6] is basically the same of the STA[9] and they differ only by the use or not of the direct summation over single particle levels to determine $A$ and $U$.

The system of equations (3.1) is non trivial because $a$ and $b$ appear in the denominators of the various terms. The relation of $a$ with the thermodynamic chemical potential, Eq. (3.3), is not given in [6], but using the steepest descent method in connection with the Darwin-Fowler

**Table I - Levels of the spherical potential well of Ref.[5]**

| level $i$ | energy $\epsilon_i$ (MeV) | level occupation | total occupation |
|---|---|---|---|
| 1s | 2.0600 | 2 | 2 |
| 1p | 4.2020 | 6 | 8 |
| 1d | 6.9420 | 10 | 18 |
| 2s | 8.2400 | 2 | 20 |
| 1f | 10.1760 | 14 | 34 |
| 2p | 12.4420 | 6 | 40 |
| 1g | 13.9670 | 18 | 58 |
| 2d | 17.2630 | 10 | 68 |
| 1h | 18.2930 | 22 | 90 |
| 3s | 18.6220 | 2 | 92 |

approximation,[3, 21] Ref.[6] considers the following expression for the nuclear level density

$$\rho(A, E) = \mathrm{e}^R a^A b^E / 2\pi (GC - B^2)^{1/2} \ , \tag{3.4}$$

where $R$ is given by

$$R = \sum_{(i)} g_i \log\left(1 + a^{-1} b^{-\epsilon_i}\right) \ , \tag{3.5}$$

and $G$, $B$, $C$ are given by

$$G = \sum_{(i)} g_i ab^{\epsilon_i} (1 + ab^{\epsilon_i})^{-2} \ , \tag{3.6}$$

$$B = \sum_{(i)} g_i ab^{\epsilon_i} \epsilon_i (1 + ab^{\epsilon_i})^{-2} \ , \tag{3.7}$$

$$C = \sum_{(i)} g_i ab^{\epsilon_i} \epsilon_i{}^2 (1 + ab^{\epsilon_i})^{-2} \ , \tag{3.8}$$

therefore, $R$ is the thermodynamic potential,[17] the logarithm of the partition function given by (2.9) and (2.14), with only the energy ($y$) and the number of particles ($x$) being considered.

Equation (3.4) results from the assumption that $G,B,C \gg 1$, but the exact meaning of this assumption must be determined by direct numerical computation. Ref.[6] obtains $G$ with magnitude close to "2" and energies not greater than 10 MeV, while our calculations for the same Model Space, for atomic masses between 13 and 79 and, give $G$ between zero and "10", increasing with $A$, for excitation varying from zero to the *point of maximum of* $G$, while $B$ is nearly one order of magnitude greater than $G$, for all excitations and mass numbers.

On the other hand, the definitions of the nuclear level density compared in the analysis of Ref.[6] are either Eq.(3.4) or the "asymptotic" nuclear density given by the statistical estimate of Bethe[9]

$$\rho(U, T') = \frac{\exp(2U/\kappa T')}{48U} \ , \tag{3.9}$$

corresponding to a nuclear temperature, $T'$,

$$\kappa T' = \frac{2}{\pi} \sqrt{\frac{\mu U}{A}} \ , \tag{3.10}$$

calculated using Eqs.(4a), (9) and (29) of Ref.[9]. The nuclear excited state is defined in an average way, using the single particle density as a continuous probability distribution to obtain the properties of the total system. The OCN temperature, $T$, is defined from (3.2) as a function of $b$, which is determined numerically as the solution of (3.1).

If the model of Ref.[6] was a consistent formulation, these two definitions should give similar results for the nuclear density if the sums over single particle states are replaced by integrals and "$a$" is very large. On the other hand, direct calculation shows that the region of values of $U$ in which $a$ increases quickly and becomes much greater than 1 corresponds to values of $b$ less than 1, which are physically meaningless. Therefore, the replacement of sums over single particle states by integrals in this model can be considered as meaningless too, or at least contradictory, and the comparison with the STA model of Ref.[9] is, strictly speaking, not possible.

The range of values of $U$ for which $a$ is greater than 1, is not physically meaningful because it corresponds to a negative chemical potential (the internal energy would *decrease* for increasing number of particles in the system), which can only be accepted if some kind of "radiative process" is included. In this case the system would be "open" with respect to the *type* of its constituent subsystems (for example, to include "fermions" *and* "bosons") and the grand canonical ensemble would have to consider this additional field, besides the nucleon field, to give a complete description of the total system. In addition, in this region of $U$ $\beta$ becomes very close to zero (very large temperatures), indicating a kind of "saturation" of the system for these excitations and higher. For higher $U$ the temperature would also become negative, corresponding to $b$ smaller than 1, which is unacceptable in terms of the usual definition of temperature, as a measure of the average kinetic energy of the component subsystems. At this "saturation point" the occupation probabilities of all single particle states would be 50%, which can be physically interpreted as meaning that *all* single particle states of the Model Space would be partially occupied with the same probability!

The meaningful region, where $b$ is greater than 1, give $a$ smaller than 1 and very small for low $U$. Therefore, in the region of low $U$ that is where the model is physically meaningful, but *theoretically inconsistent*, one should expect both definitions, (3.4) and (3.9), to yield considerable different results. In fact, the calculation shows that, depending on the model used for the description of single particle states, they *may be not very different*, especially for low $A$ or at the closed shell values. On the other hand, the two densities tend to diverge more pronouncedly for low excitations, especially the derivative of the density, for *all A*.

We interpret this similarity of results with the STA as another case of theoretical "coinci-dence", despite the cited inconsistencies of method used by Ref.[6], and the important differences at low temperatures are indicative of these inconsistencies. If the use of the DFM was coherently implemented in the OCN, the differences could be interpreted as indicating improvements for the STA estimate.

If a simplified single particle level model with *constant interspacing* between levels is used, it is possible to reproduce the results of the spherical box model very closely, by adopting an energy spacing close to the first single particle level of the box model. For smaller interspacing the OCN density, $\rho(A,E)$, tend to be larger than STA, $\rho(U,T')$, while for larger interspacing it may be always smaller than STA. Therefore, the important aspect of $\rho(A,E)$, as noted by Ref.[6], is its *oscillation* when $A$ approaches and goes away from closed shells, which is physically more realistic than the monotonic increase of $\rho(U,T')$ for increasing $A$.

Figure 1 shows "$a$" as a function of $A$ and $U$. Notice how "$a$" crosses the plane corresponding to "$a$=1" for increasing excitations as $A$ increases (although it is not very clear in the graph, for fixed values of $U$ "$a$" decreases steadly for increasing $A$), corresponding to increasing chemical potential ($\mu$) for greater $A$ and smaller $U$, which is the physically expected behavior.

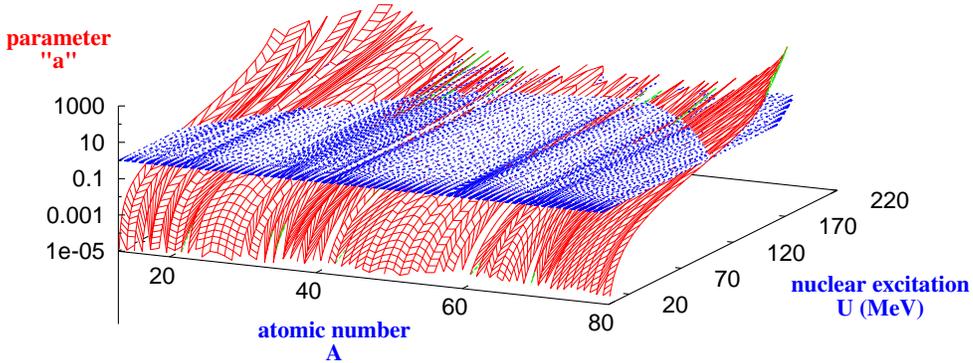

**Figure 1**. Parameter "$a$" becomes very large for large excitations but increases slower for large $A$. The physically meaningful region corresponds to excitations for which $a$ is less than 1, which is shown in the figure as the part of the graph below the plane parallel to the $xy$-plane cutting the $z$-axis at "1".

Parameter $a$ increases steadly as a function of $U$ and $b$ decreases steadly. The analogous graph for "$b$" is essentially the reverse of "$a$", with $b$ crossing the "$b=1$" plane in the downward direction at basically the same points $(A, U)$.

For very high $A$ the results are not meaningful due to the limitation imposed in the definition of the Model Space on the number of single particle levels available for transitions.

Parameter $G$ in (3.6) has an absolute maximum as a function of $U$ that depends on the atomic mass and the Model Space. This point usually corresponds to very high excitations (for a stable nuclear system) possibly reaching more than 100 MeV. The results for the spherical box model are plotted in Fig. 2 and show a not very large $G$ for low excitations. Parameter $B$ has similar behavior and reaches maximum for $U$ a little higher than the maximum of $G$.

The maximum of $G$ corresponds to excitations for which $b$ is close to 1, and slightly greater, and this region also contains the point where $a$ becomes greater than 1.

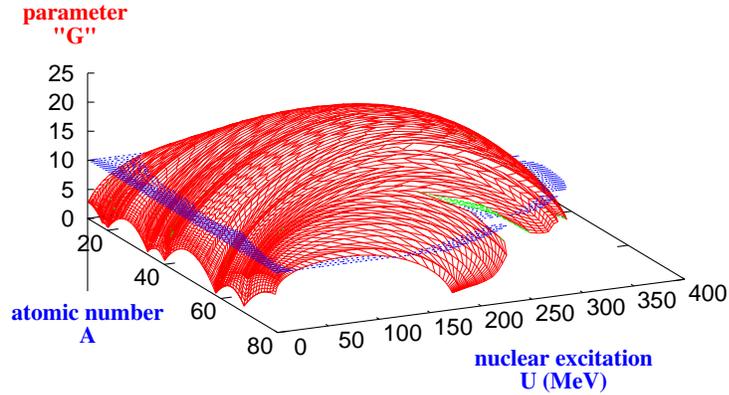

**Figure 2**. Parameter $G$, of Eq.(3.6), has an absolute maximum for very large excitations. For $U < 10$ MeV $G$ is less than 5, in agreement with Ref.[6]. Parameter $B$ has a similar functional behavior and larger magnitude.

The results of [6] for $G$ were always close to "2" because the regions of $U$ and $A$ considered were relatively limited. Our calculations show that only for $A$ close to the closed shell values it is possible to obtain $G \approx 2$ for $U$ usually smaller than 5 MeV, while for $A$ in between closed shells $G$ is always greater than 3 even for very small $U$.

Notice that these results for $G$ and $B$, in the physically meaningful region of $U$, are incompatible with the saddle-point approximation and the Darwin-Fowler method. Margenau[6] noticed the incompatibility, but did not consider it in detail.

Therefore, in the OCN the physically acceptable region is that of not very large excitations,

corresponding to small $a$, large $b$, $G$ not much greater than 1 and not very large temperatures, $T$.

In general, our calculations for the nuclear density of Eq.(3.4) compare well with Ref.[6]. We obtain an oscillation of $\rho(A,E)$ as a function of $A$, increasing when $A$ departs from the closed shell values and reaching a local minima at these values, which is also the conclusion of Ref.[6] and correctly implies maximum absorption of slow neutrons for nuclei with values of $A$ in between the closed shell values.

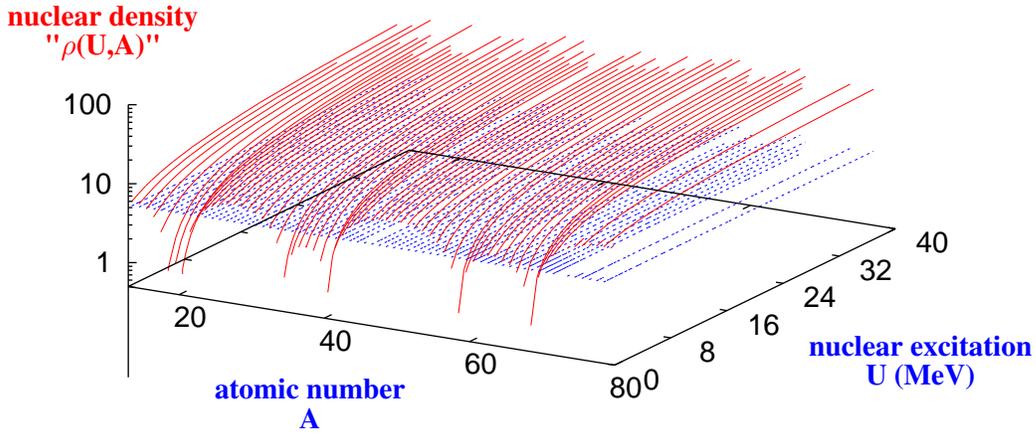

**Figure 3**. Logarithm of the density $\rho(A,E)$ of Eq.(3.4) as a function of $A$ and the nuclear excitation, oscillates and reaches local minima for closed shell values of $A$.

## 3.1 The occupation probabilities of single particle levels.

The analysis of Margenau [6] reviewed in the last section centers the definition of the occupation numbers of the single particle levels on the solution of equations (3.1) for given $A$ and $E$.

On the other hand, the direct distribution of nucleons into the available single particle states of the Model Space defines the IPMC and the level occupations as *discrete functions* of the nuclear excitation.

The occupation probability of level $n$ in Eq.(3.1), is given by

$$p(U,n) = \frac{1}{1 + ab^{\epsilon_n}},$$ (3.11)

and varies *continuously* with $U$ (see Fig. 4) and shows a smooth transition from large occupations of low energy levels for small excitations to increasing occupations of high energy levels for higher $U$.

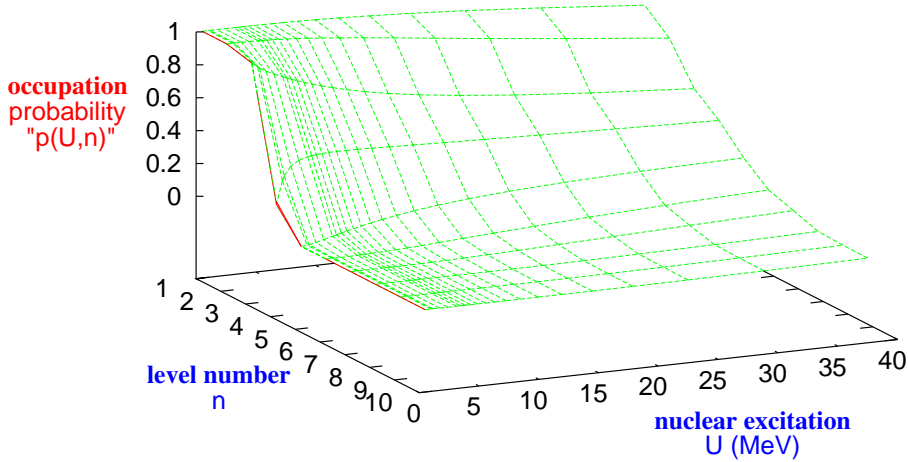

**Figure 4**. Probability of level $n$, $p(U,n)$, of the OCN description as a function of the nuclear excitation and the level $n$.

This does not necessarily happen in the IPMC occupation in which levels with zero probability may appear in between two others with non null occupation, as we see in Fig. 5.

Therefore, the two distributions are *very different* and correspond to essentially different concepts of the nuclear ensemble. The OCN description can be considered as semi-classical in the sense that it gives the possible nuclear excitations as a continuous function of the nuclear temperature, while the IPMC produces a set of discrete values of $U$, which depends on the model adopted for the single particle levels and the number of "mobile particles" in the Model Space. In fact, by definition, any excitation energy is possible in the OCN, by adjusting the parameters $a$ and $b$ in Eq.(3.1), while in the IPMC only the sums of the energies resulting from the distribution of the nucleons on the presumed scheme of single particle levels are possible and the probability distributions may oscillate in a non well-defined way, from single particle level to the next. In the OCN the probability distributions of single particle levels are continuous functions of the single particle energies.

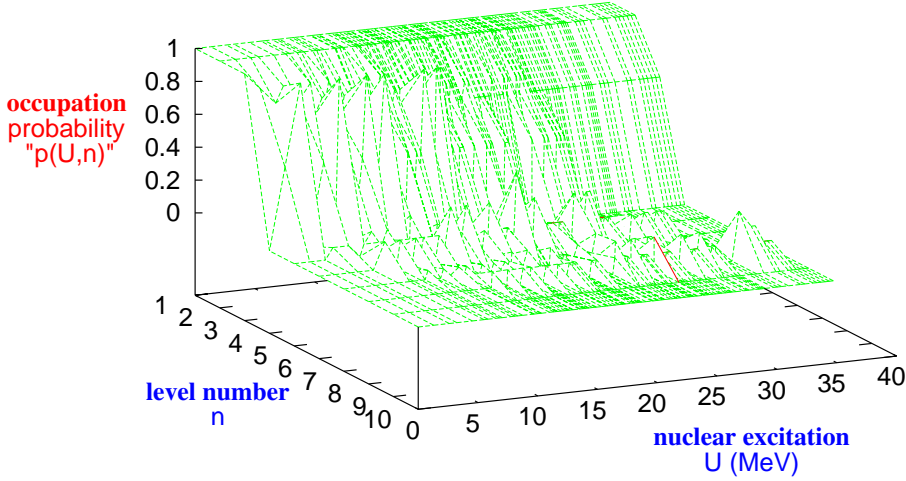

**Figure 5**. Probability of level $n$, $p(U,n)$, of the IPMC as a function of the nuclear excitation and the level $n$.

To compare the OCN and IPMC results one does not need to have all particles of the Model Space "mobile", because at least for small excitations the occupation of the first levels will be "1" or very close to "1" in both approaches, therefore the single particle states involved in the definition of the excited configuration will always be a few ones close to the Fermi level, revealing that the difference between the two approaches has an essential nature and it is not just a matter of the degree of approximation adopted.

Having in sight these important differences and for the sake of our present discussion, one may ask which approach should be considered more physically correct.

The OCN occupation should be "statistically correct", in the sense that if the number of configurations populating each nuclear level increased infinitely the corresponding level occupation probability should tend to the statistical value. There is a problem though with respect to this "interpretation" because the number of nuclear levels is *essentially* different in the definitions of the OCN (infinite) and of the IPMC (finite). In this regard and taking into account the greater consistency of the IPMC with the quantum description, the OCN must be considered as *at most* an interesting simpler approximation to the IPMC.

Notice that if (3.11) is the definition of the occupation probability and the connection of $U$ and the nuclear temperature is Eq.(3.1), then, having in sight that $U$ in the IPMC is discrete, it results that the nuclear temperature will also be discrete in this approach. Similarly, because $a$ and $b$ are univocally determined by (3.1), the chemical potential ($\mu$) is also a discrete quantity in the IPMC description.

The Fermi statistics should be a direct result of the OCN if consistent definitions are used. In this case, the two models, OCN and IPMC, should be comparable to a good extent, even if not completely identical.

## 4.   Final comments and conclusion

This work intended to give further elements of criticism to those presented on Ref.[1, 4] regarding the application of the Darwin-Fowler method (DFM) in connection with the statistical IPM approach. We focused our analysis in the theoretical consistency of various models and how they compare with the results of the IPMC.

We centered our attention on the works of Bethe[9], usually called statistical IPM or STA approach, which does not use the Darwin-Fowler method, the traditional crtical analyses of Bloch[7, 17] and Margenau[6] to the STA, which consider different approaches to define the nuclear level density and use the DFM, and other works.

The statistical IPM of Bethe is theoretically consistent, although approximate, and cannot be considered as a fundamental description, but a semi-phenomenological one. Therefore, it can be used to obtain good estimates of the level density in the region of validity of the hypotheses of the Sommerfeld model, but not to criticize the theoretical consistency of other models.

Bloch[7] tried to improve upon the results of the STA by developing a formalism in which the sums over single particle levels are *not* replaced by integrals to better describe the influence of the details of the single particle levels on the nuclear level density, but assumed without further consideration the validity of the DFM, while using the hypotheses of large degeneracy of nuclear configurations and low nuclear excitation ($\beta \rightarrow \infty$).

For a typical model of single particle states, as the Harmonic Oscillator[1] or spherical box model,[6] the number of nuclear levels per MeV, as calculated by the IPMC or estimated by phenomenological models for example,[11] can be very large, even if one does not consider the spin-isospin degeneracy, but frequently this does not happen for low nuclear excitations, specially for light nuclei,[19] and the two hypotheses can be inconsistent if the Sommerfeld model[10] is used or if one intends to use $CAP$ for nuclear levels in connection with the Darwin-Fowler method.[7]

The formalism of Ref.[7, 17] is, therefore, contradictory or inconsistent by non specifying the range of $U$ for which it is valid. Bloch manages to obtain some interesting results like Bethe's expression for the level density for one type of nucleon using the saddle point method,[7, 17] and the spin distribution function of light nuclei[7], in comparison with the "exact" results of Critchfield and Oleksa[16], but due to the lack of consistency these results can only be considered as *coincidental* rather than a positive demonstration of the validity of his approach.

In this case, this problem of consistency can be solved by adopting a different approach to describe the low energy region,[8, 11] or, in a more fundamental way, by using a single consistent theorical model for the various regions of excitation, as for example the IPMC model.

A similar consistency problem occurs in the OCN approach of Margenau[6], reviewed in Sec.3. We saw that in the high excitation region the usual thermodynamic parameters of temperature and chemical potential become negative and, therefore, physically meaningless, if described by the OCN formalism.

In the region of low excitations, where the OCN is physically meaningful, but theoretically inconsistent, one observes important differences in comparison with the STA density,[9] for all $A$, indicating that part of the conclusions of Ref.[6] were again based on theoretical *coincidences* rather than well established conclusions within the presumed model.

Therefore, we see that the use of the DFM brings important problems of theoretical consistency when used in connection with the statistical IPM[7, 17, 6] or the IPMC or OCN approaches[1, 4] and should be used criteriously, specially for low excitations, or avoided altogether and replaced by an entirely consistent method, as for example the one presented in Ref.[1].

On the other hand, the OCN and IPMC distributions of occupation probabilities reveal important differences and correspond to essentially different concepts of the nuclear system. The OCN describes the nuclear excitations as a *continuous function* of the nuclear temperature, with probability distributions of single particle levels showing a continuously varying pattern, while the IPMC produces a *finite set* of discrete excitations and the probability distributions may oscillate in a non well-defined way, among single particle levels close to each other. Strictly speaking, the latter is consistent with the quantum description, while the former is not.

Then, independently of the inconsistencies noted above in the approach of Ref.[6], the IPMC is endowed with greater physical meaning and the OCN description must be considered as at most an interesting approximation to the IPMC.

Alternate combinatorial IPM approaches exist in the literature as, for example, the one proposed by Williams[5] to calculate nuclear level densities directly from single particle levels, using a systematic computation method that is analogous to the solution of the problem of the partition of integer numbers.

But one should notice that, despite being a theoretically consistent quantum mechanical description, the IPMC has a strong dependence on the model adopted for single particle states,[1, 5] which may greatly influence the nuclear level density estimate, when compared with other models especially the powerful semi-phenomenological models based on the STA.[12, 19] For example, the results of Ref.[5] show important effects on the nuclear level density in connection with the Rosenzweig effect[8] and pairing.

In summary, the present analysis has no intention to be comprehensive, but to show additional critical arguments about the use of the DFM in statistical IPM approaches. We have shown here and in our previous analysis[4] that the Darwin-Fowler method, depite being an important tool to obtain useful algebraic approximations, should be used cautiously for non very excited systems.

In these cases, it is suggested that an alternate consistent method should be used instead and the IPMC has shown important advantages over the others analyzed in this study. For not very excited nuclear systems, as for example in the pre-equilibrium stage, some options are presented in Refs. [1] and [5].

# Appendix I - The angular momentum-isospin degeneracies

We present here a detailed description of the counting of degeneracies in connection with the angular momentum-isospin symmetry.

The essential ideas are presented in the classical literature on the subject,[13, 14, 15, 16] but this description intends to be a reference for future works on nuclear level density and also serve as a practical summary as this information is usually not given in all its details at one single reference in the perused literature.

In this analysis, the counting of occupied fermion states is constrained by the exclusion principle based on a presumed complete set of single particle quantum numbers. We assume that this complete set is defined by the "kinetic" energy, $\epsilon$, and the quantum numbers for the $z$-projections of the orbital angular momentum (OAM), $\mathtt{m_l}$, spin, $\mathtt{m_s}$, as well as the isospin projection, $\mathtt{m_t}$. The values of $\mathtt{m_s}$ and $\mathtt{m_t}$ can only be $\pm 0.5$ for "nucleons"[22] and all these qantum numbers (projections) are additive, corresponding to extensive thermodynamic quantities that can be described by the grand canonical partition function defined in Sec.1.

First we need to define the set of all possible single particle states to be "filled" with a particle, let $n$ be the number of such states and $\mathtt{m}$ the number of particles that will be distributed into the $n$ states. The number of possible nuclear configurations, $\mathcal{N}(n,\mathtt{m})$, created by the distribution of $\mathtt{m}$ identical particles (protons and neutrons are considered identical nucleons in different states defined by their isospin values) into $n$ distinct single particle states is given by the binomial coefficient

$$\mathcal{N}(n,\mathtt{m}) = \binom{n}{\mathtt{m}} \tag{I.1}$$

which counts the sets of $\mathtt{m}$ integers between 1 and $n$, ordered by increasing values.

We classify the various configurations by essentially two independent variables, according to the following scheme. To each different set of $(\epsilon,\mathtt{m_l},\mathtt{m_s},\mathtt{m_t})$ we associate a pair of integers "$xy$", where the first designates the possible values of the pair of the spin-isospin projections, $(\mathtt{m_s},\mathtt{m_t})$, with $x$=1, 2, 3, 4 corresponding to $(\mathtt{m_s},\mathtt{m_t})$=(+,+), (+,-), (-,+), (-,-), respectively, as proposed by Wigner,[15] and $y$=1, 2, ... , 9 indicates the orbital angular momentum projection, $\mathtt{m_l}$, for each sp-state. For example, in the case of $l$=1, we have $y$=1, 2, 3 corresponding to $\mathtt{m_l}$=-1, 0, +1. For $l$=2, we have $y$=1, 2, 3, 4, 5 corresponding to $\mathtt{m_l}$=-2, -1, 0, +1, +2, etc. The whole scheme for the case "$l$=2" is shown in table Table II, where the value attributed to the energy ($\epsilon$=1) is arbitrary.

The spin and isospin projections, $(\mathtt{m_s},\mathtt{m_t})$, can also be used to define the "partitions" to which the various configurations belong,[14, 15] and to classify them according to their $STY$-symmetry parameters,[15] as we will show in the next section.

We are considering here the energy "$\epsilon{=}1$" fixed, but the extension to more than one energies, $\epsilon \in \{1, 2, 3, ...\}$, is straightforward, by just increasing the number of terms corresponding to these new energies in Table II. Then, the energies of all configurations will be the same, equal to $\mathtt{m}\epsilon$, and will not enter in the counting of degeneracies. The different values of the *total* OAM ($L$), spin ($S$) and isospin ($T$) introduce a splitting of this "kinetic energy" level,[16] which we will not consider here.

The other quantum numbers can be added as the eigenvalues of the angular momentum algebra,[22] but total value of their projections for each nuclear configuration is obtained by the simple sum of the component particle values. The LS-coupling is not considered, but instead a *subtraction scheme* is used to determine the degeneracies associated with each total nuclear level quantum numbers ($L$), ($S$) and ($T$). It is convenient to designate by these letters also the sums of the corresponding single particle projections, $L{=}\sum \mathtt{m_l}$, $S{=}\sum \mathtt{m_s}$ and $T{=}\sum \mathtt{m_t}$, because the subtraction scheme presented next will actually reduce the degeneracies associated with the sums to those associated with the total quantum numbers of the configurations.

**Table II - Description of single particle levels for $l{=}2$**

| level | $xy$ | label | | | | $\epsilon$ | $\mathtt{m_l}$ | $\mathtt{m_s}$ | $\mathtt{m_t}$ |
|---|---|---|---|---|---|---|---|---|---|
| 1 | 11 | 1 | +2 | + | + | 1 | +2.0 | +0.50 | +0.50 |
| 2 | 12 | 1 | +1 | + | + | 1 | +1.0 | +0.50 | +0.50 |
| 3 | 13 | 1 | 0 | + | + | 1 | 0.0 | +0.50 | +0.50 |
| 4 | 14 | 1 | -1 | + | + | 1 | -1.0 | +0.50 | +0.50 |
| 5 | 15 | 1 | -2 | + | + | 1 | -2.0 | +0.50 | +0.50 |
| 6 | 21 | 1 | +2 | + | - | 1 | +2.0 | +0.50 | -0.50 |
| 7 | 22 | 1 | +1 | + | - | 1 | +1.0 | +0.50 | -0.50 |
| 8 | 23 | 1 | 0 | + | - | 1 | 0.0 | +0.50 | -0.50 |
| 9 | 24 | 1 | -1 | + | - | 1 | -1.0 | +0.50 | -0.50 |
| 10 | 25 | 1 | -2 | + | - | 1 | -2.0 | +0.50 | -0.50 |
| 11 | 31 | 1 | +2 | - | + | 1 | +2.0 | -0.50 | +0.50 |
| 12 | 32 | 1 | +1 | - | + | 1 | +1.0 | -0.50 | +0.50 |
| 13 | 33 | 1 | 0 | - | + | 1 | 0.0 | -0.50 | +0.50 |
| 14 | 34 | 1 | -1 | - | + | 1 | -1.0 | -0.50 | +0.50 |
| 15 | 35 | 1 | -2 | - | + | 1 | -2.0 | -0.50 | +0.50 |
| 16 | 41 | 1 | +2 | - | - | 1 | +2.0 | -0.50 | -0.50 |
| 17 | 42 | 1 | +1 | - | - | 1 | +1.0 | -0.50 | -0.50 |
| 18 | 43 | 1 | 0 | - | - | 1 | 0.0 | -0.50 | -0.50 |
| 19 | 44 | 1 | -1 | - | - | 1 | -1.0 | -0.50 | -0.50 |
| 20 | 45 | 1 | -2 | - | - | 1 | -2.0 | -0.50 | -0.50 |

The idea of the subtraction scheme is based on the angular momentum algebra. For example, if a given level has OAM $L$, it has $(2L{+}1)$ projections $\mathtt{m_L}{=}$-$L$, -($L$-1), $\cdots$, ($L$-1), $L$, that would add to the same values of projections corresponding to other OAM's with quantum numbers between 0 and $(L\text{-}1)$, i. e., the sums of projections have an intrinsic redundancy and do not define univocally the total quantum number for the configuration.

For example, if for a given pair of $(S,T)$ we count 10 configurations with "$\sum \mathtt{m_l}=2$" this does not necessarily mean that all them are associated with levels with "$L=2$" ("$D$" levels), as they could also come from levels with "$L>2$".

Therefore, because we are summing over the *projections* of the angular momenta and the isospin, to obtain the degeneracy associated with a given $L$ we must subtract the sum of $\mathtt{m_l}$ corresponding to $(L)$ by the sum of $\mathtt{m_l}$ corresponding to $(L\text{+}1)$, as part of the degeneracy that we are counting for $(L)$ will be due $(L\text{+}1)$, and so on, subtract the sum corresponding to $(L\text{+}1)$ by the sum of $\mathtt{m_l}$ corresponding to $(L\text{+}2)$, etc.

The subtraction procedure must, therefore, be realized from top down, starting from the highest possible value of $\sum \mathtt{m_l}$ and going downward to the lowest, usually "0", for each set of configurations defined by given values of $\sum \mathtt{m_s}$ and $\sum \mathtt{m_t}$.

More specifically, if the top OAM of a given group of configurations, defined by fixed $(S,T)=(\sum \mathtt{m_s}, \sum \mathtt{m_t})$, is "$F$" (corresponding to nuclear OAM $L=3$) possessing degeneracy "2" (i. e., we have counted 2 nuclear configurations with $\sum \mathtt{m_l}$ equal "3") and we have counted 5 nuclear configurations with sum of $\sum \mathtt{m_l}$ equal "2" (corresponding to nuclear OAM $L=2$, or a "$D$" nuclear level), then the effective degeneracies of these two nuclear levels will remain "2" for the $F$ level, because it is the top level, and become "(5-2)=3" for the $D$ level, using the subtraction scheme.

After defining all possible configurations, one may start the separation into different levels by the values of $\sum \mathtt{m_l}$, with fixed $(\sum \mathtt{m_s}, \sum \mathtt{m_t})$. Only non negative values must be considered for $\sum \mathtt{m_l}$ because after the subtraction procedure, the remaining value for each component will coincide with the corresponding total OAM. For the spin projections, $S=\sum \mathtt{m_s}$, and isospin projections, $T=\sum \mathtt{m_t}$, we may keep all components to make it easier to visualize the degenerate states for each $l$.

Then, we count the number of times each positive values of $\sum \mathtt{m_l}$ appears, collect the corresponding set of configurations and use the subtraction scheme to define the counting corresponding to the total quantum numbers. We order these results by their values of $^{2S+1}L$ in increasing order of $(2S+1)$. Then, the degeneracy corresponding to $^{2S+1}L$ will be the subtraction of the counting for $(2S+1)$ by the counting for $(2(S+1)+1)$, for given $L$ and all $S$. These values can then be compared with the tables of the literature.[13, 14, 16, 23]

To define the degeneracy for levels with different isospins, we proceed as before but keep the total isospin, $T$, along with $L$ and $S$ in the description of the nuclear configurations. Then, for a given $L$, consider a set of configurations with OAM equal $L$, ordered by decreasing values of $(2S+1)$ and $T$, and the first of these configurations with counting greater than zero we consider as the "reference" configuration for purposes of counting, with nuclear parameters that we will designate by $L_r$, $S_r$, $T_r$. Then, we look for configurations in this set (given $L$) with $S \leq S_r$ and/or $T \leq T_r$ and greater number of configurations counted than the number of the reference configuration. We subtract the larger counting by the smaller one, to eliminate the counted states that are mere projections of the reference level. After all subtractions of a given $S$ are considered we look for the next lower value of $S$ with positive counting of degeneracies and make it the new "reference", etc. It is important to notice that *the subtracted values must be kept for the next round* of subtractions. The procedure must be repeated for all values of $S$ for the given $L$ and be repeated for all values of $L$.

For example, suppose we have the following set of degeneracies for "$L=0$" (label "$S$"), where the superscript of "$S$" represents the $(2S+1)$ spin factor.

**Table III - Typical count of degeneracies**

| Dgeneracy | level | isospin $T$ |
|-----------|-------|-------------|
| 02 | $^7$S | 0.00 |
| 01 | $^5$S | 2.00 |
| 05 | $^5$S | 1.00 |
| 08 | $^5$S | 0.00 |
| 05 | $^3$S | 2.00 |
| 12 | $^3$S | 1.00 |
| 22 | $^3$S | 0.00 |
| 02 | $^1$S | 3.00 |
| 08 | $^1$S | 2.00 |
| 22 | $^1$S | 1.00 |
| 32 | $^1$S | 0.00 |

Then, applying the subtraction rules to the set of degeneracies of Table III, yields the following final count,

**Table IV - Degeneracies after subtraction procedure**

| degeneracy | level | isospin $(T)$ | cumulative count |
|---|---|---|---|
| 02 | $^7$S | 0.00 | 14 |
| 01 | $^5$S | 2.00 | 39 |
| 04 | $^5$S | 1.00 | 99 |
| 01 | $^5$S | 0.00 | 104 |
| 04 | $^3$S | 2.00 | 164 |
| 03 | $^3$S | 1.00 | 191 |
| 07 | $^3$S | 0.00 | 212 |
| 02 | $^1$S | 3.00 | 226 |
| 01 | $^1$S | 2.00 | 231 |
| 07 | $^1$S | 1.00 | 252 |

where the column "cumulative count" represents the total cumulative number of nuclear states up to the given level. In the above example, the level ($^1S$, $T$=0) in Table III was composed only by projections of levels with higher spin and/or isospin and, therefore, it does not appear in Table IV.

The usual "labels" of the nuclear levels in the literature, are based on the *total* OAM ($L$), spin ($S$) and isospin ($T$) values for the various configurations. For example a level with label ($^4D$, $T$=1.0) is a level with "$L$=2" and "$S$=1.5", etc., where the superscript of the OAM symbol is (2$S$+1).[23]

The total number of states of all degenerated levels is the sum of all degeneracies, including isospin, and it is given by (I.1). For example, in the case of $L$=2 (see Table II) and 6 nucleons and holes at the Fermi level, it is

$$\binom{20}{6} = \frac{27907200}{720} = 38760 \ , \tag{I.2}$$

which could be enough to validate the hypotheses of the Sommerfeld model,[10, 7, 9] if the total energy of the levels created by the splitting associated with the spin-isospin symmetry are close enough to the ground state energy, $\mathtt{m}\epsilon$.

## 1.1 The STY parameters

The STY-parameters were created by Wigner[15] to describe the spin-isospin symmetry of nucleons, following the definition of Heisenberg[22], which behave essentially as one type of particle in 2 different states of intrinsic spin angular momentum, with projections $m_s=\pm1/2$, and 2 different states of another intrinsic quantum number, which obeys similar composition rule as the angular moementum and is related to the electric charge, the isospin, also with projections $m_t=\pm1/2$.

Therefore, the pair spin-isospin defines a symmetry group similar to the group of rotations, which can be totally described by 3 parameters "S", "T" and "Y" plus the total number of particles.[15, 24]

To introduce these parameters in a systematic way we may start with the idea of "partitions" as defined by Feenberg *et.al.*,[14] as the number of single particle states with a given OAM projection, $m_l$, in a given nuclear state (configuration).

For example, if we consider nuclear configurations with 3 single particle states, a partition denoted by [3] is one in which the OAM projections of of all 3 "particles" (sp-states) are the same, the partition [2+1] means that 2 particles possess one $m_l$ and the remaing particle a different projection, [1+1+1] means that each one of the particles of the configuration posseses a different $m_l$, [1+1+1+1] means a configuration of 4 nucleons in which all nucleons have different $m_l$, etc.

Therefore, as we have seen in the previous section, the complete description of nuclear configurations, in terms of all four single particle quantum numbers ($\epsilon,m_l,m_s,m_t$) or the pair of integers "$xy$", can be made in terms of a composition of the above idea of partition, which counts the number of $y$'s with the same value, plus the 4 possible states of the spin-isospin projections.

Another important concept that helps in the definition of the STY-parameters is that of "arrangements",[16] which is the number of single particle states, in a given nuclear configuration, occupied by "4", "3", "2" or "1" nucleons, classified by their OAM projections.

For example, an arrangement indicated by {1021} corresponds to 1 single particle state with 4 nucleons with OAM projection $m_{l1}$, zero single particle states with OAM projection $m_{l2}$, 2 single particle states occupied with 2 nucleons with a different OAM projection $m_{l3}$ and 1 single particle state occupied with 1 nucleon with a third different OAM projection $m_{l4}$, giving a total of $4+0+2\times2+1=9$ nucleons in the configuration.

Notice that we cannot put 4 nucleons in the same $m_l$ state unless they all have different pairs of ($m_s,m_t$), because we are supposing the $\epsilon$ and $m_l$ quantum numbers to be the same. If we also consider *hole states* in the description of nuclear configurations[16] the maximum "occupation" for nucleons+holes in each single particle state will be 8.

Using capital letters to designate the sums of the single particle projections as before and $n$ for the total number of particles in each nuclear configuration then Ref.[15] gives the following definition for the STY-parameters $S=2\times\sum m_s$, $T=2\times\sum m_t$ and $Y=4\times\sum m_s\times m_t$, where the factors "2" and "4" have been added to agree with the literature.[15, 16]

Notice that the sums in the definitions of $(S,T,Y)$ run over the changeable particle states (i. e., those that do not belong to closed shells in the ground state) and hole states of the nuclear configuration and, therefore, each set of STY-parameters is calculated for *one* nuclear configuration, although the final set can be common to many different configurations. Ref.[15] calls the set of different configurations with the same STY-parameters a "multiplet". All nuclear states belonging to a multiplet are expected to have the same energy and to define, therefore, an indepedent nuclear level, corresponding to a *fine structure splitting* of the level defined by the total kinetic energy of all nucleons in the ground state, as defined in the previous section.[15, 16]

The above definitions, the sums defining $(S,T,Y)$, correspond to the *projections* of the STY-parameters, the parameters used to calculate the potential energy and level splitting are the *maximum positive values of these projections.*

For example, let us consider the case of single particle states with "$l=2$" and 4 types of single particle states, associated with "protons" and "neutrons" *plus* their respective hole states. This corresponds in Eq. (I.2) to "$n=20$" and "$\mathtt{m}=8$" due to the spin degneracy for each single particle state.

Then, the arrangemnt $\{0,2,0,2\}$ corresponds to the partition [3+3+1+1] and the possible different sets of projections of STY-parameters are given in Table V, corresponding to the following different sets of $(S,T,Y)$:

$(S,T,Y)=(000)$, with projection $\{0,0,\ 0\}$,

$(S,T,Y)=(200)$, with projections

$\qquad \{2,0,0\}$, $\{0,2,0\}$, $\{0,0,2\}$ and $\{0,0,-2\}$,

$(S,T,Y)=(110)$, with projections

$\qquad \{1,1,0\}$, $\{1,0,1\}$, $\{1,0,-1\}$, $\{0,1,1\}$ and $\{0,1,-1\}$

$(S,T,Y)=(220)$, with projections

$\qquad \{2,2,0\}$, $\{2,0,2\}$, $\{2,0,-2\}$, $\{0,2,2\}$ and $\{0,2,-2\}$,

$(S,T,Y)=(211)$, with projections

$\qquad \{2,1,1\}$, $\{1,2,1\}$ and $\{1,1,2\}$,

and

$(S,T,Y)=(21\text{-}1)$ with projections

$\qquad \{2,1,-1\}$, $\{1,2,-1\}$ and $\{1,1,-2\}$,

which can be compared with the corresponding arrangement in Table I of Ref.[16].

**Table V - Symmetry Parameters Projections**

| S | T | Y |
|---|---|---|
| 0 | 0 | 0 |
| 0 | 0 | 2 |
| 0 | 0 | -2 |
| 0 | 1 | 1 |
| 0 | 1 | -1 |
| 0 | 2 | 0 |
| 0 | 2 | 2 |
| 0 | 2 | -2 |
| 1 | 0 | 1 |
| 1 | 0 | -1 |
| 1 | 1 | 0 |
| 1 | 1 | 2 |
| 1 | 1 | -2 |
| 1 | 2 | 1 |
| 1 | 2 | -1 |
| 2 | 0 | 0 |
| 2 | 0 | 2 |
| 2 | 0 | -2 |
| 2 | 1 | 1 |
| 2 | 1 | -1 |
| 2 | 2 | 0 |

# References

[1] F. B. Guimaraes and B. V. Carlson, *A direct microscopic approach to transition strengths in pre-equilibrium reactions* (2011), arXiv:1106.4283v2 [nucl-th].

[2] B. V. Carlson, personal communication containing the original ideas of a microscopic formalism for transition strengths in nuclear pre-equilibrium and the original version of code TRANSNU (2005).

[3] C. G. Darwin and R. H. Fowler, Phil.Mag. **44**, 450 (1922); C. G. Darwin and R. H. Fowler, Phil. Mag. **44**, 823 (1922).

[4] F. B. Guimaraes, *Brief crtical analysis of the Darwin-Fowler method* (2011). arXiv:1109.1164v1 [nucl-th].

[5] F. C. Williams, Jr., Nucl. Phys. **A 133** , 33 (1969).

[6] H. Margenau, Phys. Rev. **59** , 627 (1941).

[7] C. Bloch, Phys. Rev. **93**, 1094 (1954).

[8] N. Rosenzweig, Phys. Rev. **108** , 817 (1957).

[9] H. A. Bethe, Phys. Rev. **50**, 332 (1936).

[10] A. Sommerfeld, Zeits. f. Physik **47**, 1 (1928).

[11] A. Gilbert and A. G. W. Cameron, *Can. J. Phys.*, **43**, (1965) 1446.

[12] A. Mengoni and Y. Nakajima, *J. Nucl. Sci. Tech.*,**31** no.2, 151 (1994); JAERI-M 93-177, Japan Atomic Energy Research Institute, Sep. 1993.

[13] H. A. Bethe and R. F. Bacher, Rev. Mod. Phys **8**, 82 (1936).

[14] E. Feenberg and E. Wigner, Phys. Rev. **51**, 95 (1937);
E. Feenberg and M. Phillips, Phys. Rev. **51**, 597 (1937).

[15] E. Wigner, Phys. Rev. **51**, 106 (1937).

[16] C. L. Critchfield and S. Oleksa, Phys. Rev. **82**, 243 (1951).

[17] C. Bloch in *Les Houches Lectures (1968)*, pg. 305, ed. C. De Witt and V. Gillet (Gordon & Breach, New York, 1969).

[18] F. C. Williams, Jr., Nucl. Phys. **A 166** , 231 (1971).

[19] C. Y. Fu, *A Consistent Nuclear Model For Compound and Precompound Reactions with Conservation of Angular Momentum*, Report ORNL/TM-7042 (1980), Oak Ridge National Laboratory, U.S.A.;
K. Shibata and C. Y. Fu, Recent Improvements of the TNG Statistical Model Code, ORNL/TM-10093, (August 1986); C. Y. Fu, *Nucl. Sci. Eng.* **92**, 440 (1986);
C. Y. Fu, *Nucl. Sci. Eng.* **100**, 61 (1988); C. Y. Fu, *Nucl. Sci. Eng.* **86**, 344 (1984);
F. B. Guimaraes and C. Y. Fu, *TNG-GENOA User's Manual*, Technical ReportORNL/TM-2000/252 (2000), Oak Ridge National Laboratory, U.S.A..

[20] A.L.Fetter and J.D.Walecka - *Quantum Theory of Many-Particle Systems* (Ed. McGraw Hill, New York, 1971).

[21] C. van Lier and G. E. Uhlenbeck, Physica **4** , 531 (1937).

[22] W. Heisenberg, Zeits. f. Physik **77**, 1 (1932).

[23] I. E. McCarthy, *Introduction to Nuclear Theory*, Ed. John Wiley & Sons, Inc., New York (1968).

[24] J. A. de Wet, Proc. Camb. Phil. Soc. (1971) **70**, 485.